\documentclass[twocolumn,nobibnotes,showpacs,preprintnumbers,amsmath,amssymb]{revtex4}
\pdfoutput=1
\usepackage{graphicx, hyperref}
\usepackage{dcolumn}
\usepackage{amsmath}
\usepackage{bm}

\newcommand{\be}{\begin{equation}}
\newcommand{\ee}{\end{equation}}
\newcommand{\bea}{\begin{eqnarray}}
\newcommand{\eea}{\end{eqnarray}}

\begin{document}
\title{A Simple $U(1)$ Gauge Theory Explanation of the Diphoton Excess}
\author{Spencer Chang}

\affiliation{Department of Physics and Institute of Theoretical Science, University of Oregon, Eugene, Oregon 97403}

\begin{abstract}
The recent ATLAS and CMS diphoton resonance excesses are explored in a simple $U(1)$ gauge theory extension of the Standard Model where the resonance is the Higgs boson of the $U(1)$ symmetry breaking, $\phi$.  This particle couples to exotic quarks which, through loops, can produce a large enough rate to explain the excess.  Due to the choice of $U(1)$ charges, flavor constraints are naturally suppressed, allowing arbitrary flavor violation in the decays of the new quarks to up-type quarks, modifying their signal topologies.  An additional heavy quark in the model decays to the lighter exotic quark by emitting either $\phi$ or the $U(1)$ gauge boson, $A_x$, giving extra signals containing diphoton and digluon resonances.  Finally, the new Higgs can decay into $\gamma  A_x$ and $Z A_x$, followed by $A_x$ decaying into Standard Model fermions through kinetic mixing.  Thus, this model gives interesting modified signals to the general class of exotic quark models explaining the diphoton resonance.  

\end{abstract}
\maketitle
\section{Introduction}
ATLAS and CMS using their early run 2 data have recently reported an interesting diphoton excess at a mass near 750 GeV \cite{CMSATLAS13}.  Coupled with a smaller excess in the CMS 8 TeV analysis \cite{CMS8}, these results have led to a tantalizing possibility  that a new particle is being discovered, potentially the first sign of new physics beyond the Standard Model to be discovered at the LHC.      

There are many interesting theoretical frameworks being considered for the diphoton excess \cite{everybody}, many focusing on a new spin 0 boson $\phi$.  This particle can fit the observed excess if it couples to new quarks, $\chi$, inducing decays into gluons and photons, which  allows it to be produced by gluon fusion.  The interactions beyond gauge interactions are
\bea
{\cal L}_{new} = -(M_{\chi}+\lambda \phi) \chi \chi^c + h.c.
\eea 
In addition, one must add an operator to allow $\chi$ to decay so as to avoid constraints from heavy stable charged particles.  Interestingly, the diphoton rate is large enough that a minimal sector of a single pair of new quarks tends to  lead to constraints from direct searches.  For instance, if the new quark is a top quark partner, constraints require it to be heavier than roughly 800 GeV.  At the same time, to fit the diphoton excess, one requires  $M/\lambda$ to be smaller than the weak scale to fit the diphoton rate, leading to  a nonperturbative strength for $\lambda$.  This suggests either a new strong group enhancing this coupling or exotic gauge representations for the quarks under the electroweak group \cite{everybody}.  In the latter proposal, the larger electric charges can achieve enhanced diphoton rates and avoid nonperturbative couplings.


As a specific example, new quarks transforming as a $SU(2)$ doublet with hypercharge $7/6$ contain a charge 2/3 quark and a charge 5/3 quark.  In this case,  the diphoton excess requires $M_\chi/\lambda < 1100\; \text{GeV}$, leading to perturbative values of $\lambda$ .  Moreover,  these quarks can decay through the interaction 
\bea
\kappa\; \chi H t^c + h.c.
\eea
where by considerations from flavor constraints we expect that it couples primarily with the right-handed top quark $t^c$. This leads to decays of the charged 5/3 quark into $W^+ t$ and the charge 2/3 quark into $Z t, \; h t$.  Such new quark decays are  actively being searched with constraints on the charge 5/3 quark requiring it to be heavier than 950 GeV \cite{CMSfivethirds}.  So in this simple model, we are allowed perturbative values of $\lambda$ to realize the diphoton excess and still be consistent with current direct searches.

Since this minimal approach is so effective in generating the excess, it is important to consider how robust the correlated signals from the new quarks are.  The diphoton excess will lead to  renewed interest in their searches, so it is important to check if their phenomenology can vary.  Another unresolved issue in these models is a motivation for the existence of $\phi$ and the assumed couplings for the new quarks.  As an investigation in this direction, we consider the possibility that $\phi$ is the Higgs excitation of a new vacuum expectation value $f$.  The simplest possibility is that $f$ breaks a global discrete symmetry, but here we choose a gauged symmetry as the resulting heavy gauge boson adds interesting new signals beyond the minimal model. Specifically, we will introduce a new $U(1)$ gauge symmetry with two  sets of new quarks.  As we will see, by a clever choice of charges for the quarks, flavor issues are avoided and thus the quarks can now decay into any of the three flavors of  up-type quarks.  In addition, the $\phi$ particle can decay into the heavy $U(1)$ gauge boson and the quark decays can produce $\phi$ particles allowing new signals in which to study these  particles.  Thus, this simple model gives a plethora of correlated new signals that can help distinguish this from other realizations of the diphoton excess.   

The outline of the rest of the paper is as follows.  In Sec.~\ref{sec:model}, we describe in detail the particle content and interactions of our model.  In Sec.~\ref{sec:phi}, we analyze the $\phi$ phenomenology, showing that we can fit the diphoton resonance and pointing out  additional decays into the $U(1)$ sector.  In Sec.~\ref{sec:quarks}, the quark phenomenology is discussed in detail.  Finally in Sec.~\ref{sec:conclusions}, we conclude.

\section{Model \label{sec:model}}
Our model consists of taking  the Standard Model and adding a $U(1)_x$ gauge boson $A_x$, a Higgs field $\Sigma$ to break the symmetry, and  two new vector-like fermions  with the gauge representations as shown in Table \ref{table:particlecontent}.  
\begin{table}[htpb]
\begin{tabular}{c|c|c|c|c}
Field & $SU(3)_c$ & $SU(2)_L$ & $U(1)_Y$ & $U(1)_x$ \\
\hline
$\Sigma$ & 1 & 1 & 0 & 1 \\
$\chi_1$ & 3 & 2 & 7/6 & 1 \\
$\chi_1^c$ & $\bar{3}$ & 2 & -7/6 & -1 \\
$\chi_2$ & 3 & 2 & 7/6 & 2 \\
$\chi_2^c$ & $\bar{3}$ & 2 & -7/6 & -2 \\ \hline
\end{tabular}
\caption{\label{table:particlecontent} Particle content of the model}
\end{table}
This enables the following interaction Lagrangian to be written down
\bea
{\cal L}_{int}&=& - M_1 \chi_1 \chi_1^c - M_2 \chi_2 \chi_2^c - \lambda_1 \Sigma \chi_1 \chi_2^c - \lambda_2 \Sigma^* \chi_2 \chi_1^c \nonumber
 \\ &&- \frac{1}{\Lambda} \Sigma^* H \chi_1 u^c  + h.c. \label{eq:couplings}\eea
In addition, we can write down a potential for $\Sigma$ such that it breaks the $U(1)_x$ symmetry with a vev $f$.  Our  normalization is where $\Sigma = (f+ \phi)/\sqrt{2}$, giving $A_x$ a mass $g_{x} f$.  In order not to dominate the $\phi$ decays with tree level decays into $A_x$ pairs, we need the $A_x$ mass to be greater than $m_\phi/2 = 375$ GeV, requiring $g_x > 0.375\; (\text{TeV}/f)$.  Given that the  the exotic quarks are charged under both $U(1)_Y, U(1)_x$, kinetic mixing occurs with size
\bea
\epsilon \sim \frac{e g_x N_c}{16\pi^2}
\eea
leading to $A_x$ decaying into Standard Model fermions proportional to their charge.  

The radial mode $\phi$ has a mass, $m_\phi^2 = 2\lambda_{\Sigma} f^2$ which we fix to  750 GeV to fit the diphoton resonance.  We assume that the potential has a  suppressed $|\Sigma|^2 |H|^2$ quartic coupling, since this leads to mixings between $h, \phi$ which can negatively affect their phenomenology.  Such mixed couplings are sufficiently small if they solely arise due to renormalization group running.  We've chosen the hypercharge of the exotic new quarks so as to increase the rate for diphotons   while also  allowing a decay for the lightest state through the dimension five operator (for an overview of such exotic quarks, see \cite{Aguilar-Saavedra:2013qpa}).  Since this operator is  suppressed by a large scale $\Lambda$, it is naturally small and thus easy to avoid any issues with flavor.  So we are now allowed to consider a general linear combination of up-type quarks in $u^c$, allowing a broader set of decays than into just top quarks.  We ignored a dimension 6 operator allowing the $\chi_2$ to decay since for a large enough scale $\Lambda$, it will be subdominant to decays induced by mixing into the dimension 5 operator.   

The mass matrix for the new fermions can be diagonalized, where the heavier, lighter  (H, L respectively) mass eigenstates are 
\bea
\chi_H &= \sin \theta\, \chi_1 + \cos \theta\, \chi_2,\; \chi_L &= \cos \theta\, \chi_1 - \sin \theta\, \chi_2,
\nonumber \\[-.2cm]\\ 
\chi_H^c &= \cos \theta^c\, \chi^c_1 + \sin \theta^c\, \chi^c_2,\; \chi_L^c &= -\sin \theta^c\, \chi_1^c +\cos \theta^c\, \chi_2^c \nonumber
\eea
with a degenerate pair of charge 2/3 and 5/3 quarks at two different masses.
These can be solved for in general, but we'll take an illustrative limit is where  $M_1=0$ and  $M_2\ll \lambda_1 f, \lambda_2 f$, leading to small mixing angles.  In this limit, we find
\bea
M_2\; \rm{ small:}&& \sin \theta \to \frac{\sqrt{2} \lambda_1 M_2}{f(\lambda_2^2-\lambda_1^2)},\quad  \sin \theta^c \to \frac{\sqrt{2} \lambda_2 M_2}{f(\lambda_2^2-\lambda_1^2)}\nonumber \\
&&m_L \to  \frac{\lambda_1 f}{\sqrt{2}}- \frac{\lambda_1 M_2^2}{\sqrt{2}f(\lambda_2^2-\lambda_1^2)} \label{eq:limmixmass}\\
&&m_H \to  \frac{\lambda_2 f}{\sqrt{2}}+ \frac{\lambda_2 M_2^2}{\sqrt{2}f(\lambda_2^2-\lambda_1^2)}\nonumber 
\eea
where in this expression, we've also assumed $\lambda_2> \lambda_1$.  The diphoton signal can be fit with $M_2$ set to zero, but as we will see, a nonzero value allows novel decay phenomenology for the heavier of the new quarks.   

 This diagonalization allows us to write down the interactions to $\phi$ in the mass basis, 
\bea
{\cal L}_{\phi}&=& - \phi(c_{\phi LL}\; \chi_L \chi_L^c   - c_{\phi LH}\; \chi_L \chi_H^c \\&& - \nonumber c_{\phi HL} \; \chi_H \chi_L^c - c_{\phi HH} \; \chi_H \chi_H^c)+h.c.
 \eea  
Expressions for these couplings in the small $M_2$ limit are 
\bea
c_{\phi LL}&\to& \frac{\lambda_1}{\sqrt{2}}+ \frac{\lambda_1 M_2^2}{\sqrt{2}f^2(\lambda_2^2-\lambda_1^2)}\nonumber \\
c_{\phi LH}& \to& 0\nonumber\\[-.2cm] \\
c_{\phi HL} &\to& -M_2/f\nonumber \\
c_{\phi HH}&\to& \frac{\lambda_2}{\sqrt{2}}- \frac{\lambda_2 M_2^2}{\sqrt{2}f^2(\lambda_2^2-\lambda_1^2)}\nonumber
\eea
and with these couplings determined, we can now analyze the phenomenology.
\section{$\phi$ decays \label{sec:phi}}
Given the exotic quark couplings to the $\phi$ particles, it is straightforward to work out the decay widths into photons and gluons, determining the signal rates for the 750 GeV resonance.  As a proxy for what we should be fitting for, it is important to look at the CMS, ATLAS analyses at 8 \cite{CMS8,ATLAS8} and 13 TeV \cite{CMSATLAS13}.  The ATLAS 13 TeV diphoton analysis has the largest excess with 8+4 excess events in two neighboring 40 GeV bins.  In our model, the width of $\phi$ is narrow, so that we expect a single 40 GeV window to contain all of our signal events.  Moreover, the ATLAS 13 and 8 TeV analyses are in mild tension (2.2 $\sigma$ for a narrow width resonance \cite{CMSATLAS13}).  Thus, in our model, it is more likely that the ATLAS 13 TeV analysis has an upward fluctuation and thus we consider a simple average of the two 40 GeV bins (i.e. 6 events) to be a reasonable estimate for the number of 13 TeV signal events at ATLAS we should fit to.  At the same time, the CMS 13 TeV analysis has 5 excess events in two neighboring 20 GeV bins.  Given the relative luminosities of 3.2 and 2.6 fb$^{-1}$ respectively, these two numbers are roughly consistent with each other. 

 A simple likelihood analysis for CMS, given 9 observed events and 4.1 expected background events, gives a 1$\sigma$ range of  $[1.9,7.9]$ signal events at CMS.  After unfolding by a $40\%$ efficiency (we take the ATLAS efficiency \cite{CMSATLAS13} since only ATLAS analyzes a spin 0 resonance) and the integrated luminosity of 2.6 fb$^{-1}$, gives a 1$\sigma$ range of $[1.8,7.6]$ fb for $\sigma(pp \to \phi \to \gamma\gamma)$.  Since there is a lower number of events at CMS, this range is larger than  the range that we would infer from the ATLAS analysis, so we'll take this as our estimate of the signal cross section our model should aim for.  
 
\begin{figure}
\includegraphics[scale=0.7]{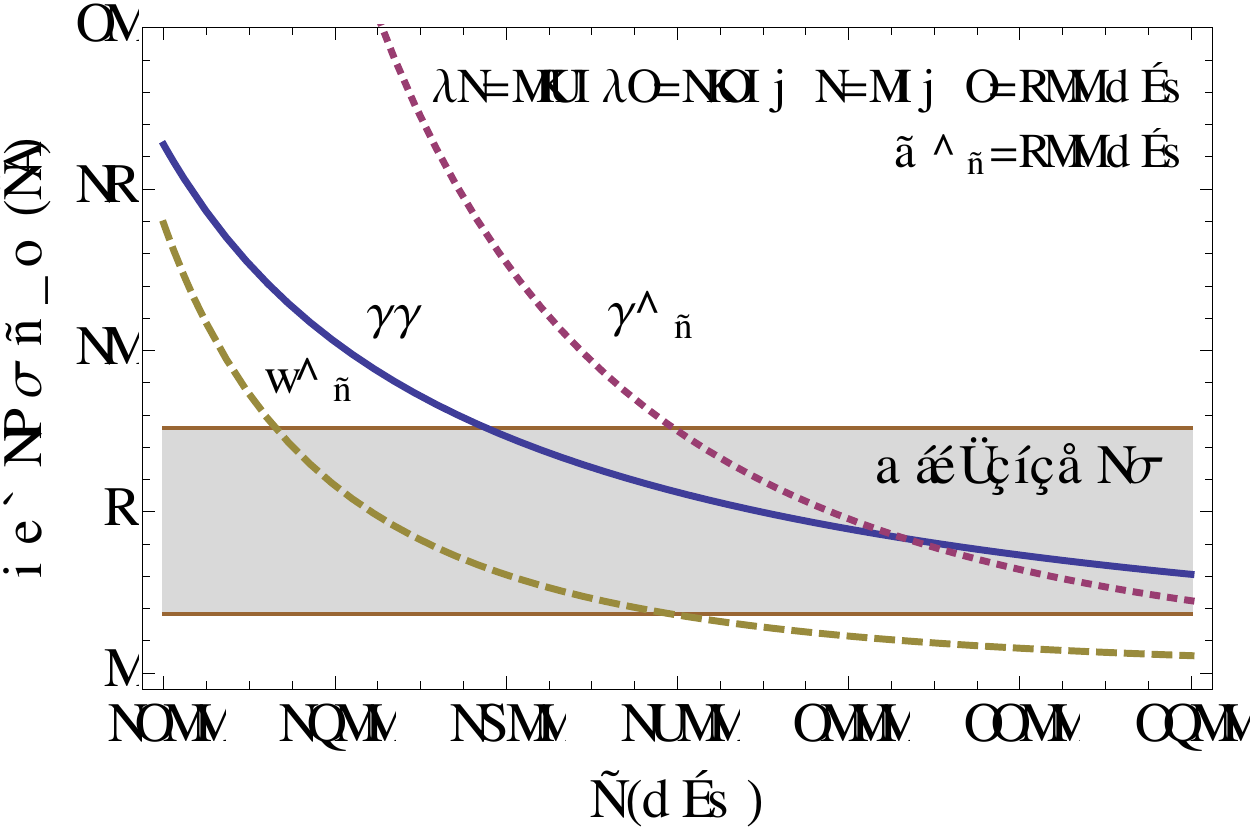}
\caption{\label{fig:diphotons}
Rates for the $\phi$ decay to diphotons (solid), $\gamma A_x$ (dotted), and $Z A_x$ (dashed) at 13 TeV LHC with the listed benchmark values.  The shaded region is our estimate for the 1$\sigma$ preferred region for the diphoton rate, using the 13 TeV analyses \cite{CMSATLAS13}.}
\end{figure}

As a benchmark, we choose the values $\lambda_1=0.8, \lambda_2=1.2, M_2=500$ GeV and plot the diphoton signal rate as a function of $f$ in Fig.~\ref{fig:diphotons}.  These $O(1)$ values of the couplings combined with the large charges of the new quarks allow us to easily fit our $1\sigma$ signal region (shaded region) for f $>$ 1600 GeV.  At lower values of f, we can get a substantially higher rate for diphotons, which can either be lowered by additional decays for $\phi$ or could be preferred if our signal region is conservatively too low.  Note that as stated earlier, we are forbidding the possibility of $\phi \to A_x A_x$ decays, so in this benchmark we fix $m_{A_x} = 500$ GeV.  Given the smallness of the kinetic mixing, higher order processes like $\phi \to A_x A_x^*$, where one of the gauge bosons is off shell, should be negligible.

Additional signals for the $\phi$ resonance include decays to gluons and $Z\gamma$ as pointed out in recent work \cite{everybody}.  In terms of these Standard Model modes, we find the rate ratios
\bea
\gamma\gamma : g g : WW : ZZ: \gamma Z= 1: 14 : 0.85 : 0.54 : 0.06
\eea
to be essentially independent of $f$.  These rates are consistent with existing constraints, but with more data and analysis improvements could become interesting.   
In this model, there are also  loop induced decays into $\gamma A_x$ and $Z A_x$ whose signal rates are plotted in Fig.~\ref{fig:diphotons}.  As the mass of $A_x$ is increased in our benchmark, these additional decays become suppressed due to phase space, eventually going to zero.  These mixed decays  lead to difermion resonances from $A_x$ (often into charged leptons) which paired with the photon or $Z$ would reconstruct to 750 GeV.  Thus,  these events could already be in existing data and would give a striking signal correlated with the diphoton excess.  

\section{Exotic Quarks \label{sec:quarks}}
The exotic quarks in this model have their own interesting phenomenology.  Many of the diphoton papers \cite{everybody} have included such quarks.  The novel nature of our quarks is that the $U(1)_x$ charge constrains the mixings and decays to Standard Model quarks to a dimension five operator.  This means that flavor constraints are easily avoided and thus the up-type quarks it decays into do not have to dominantly go into the top quark.  Furthermore, the $U(1)_x$ charges in our model lead to fermion cascades that can produce $\phi$ and $A_x$ giving distinctive signatures that can be used to distinguish this scenario.    

\begin{figure}
\vspace{.3cm}
\includegraphics[scale=0.7]{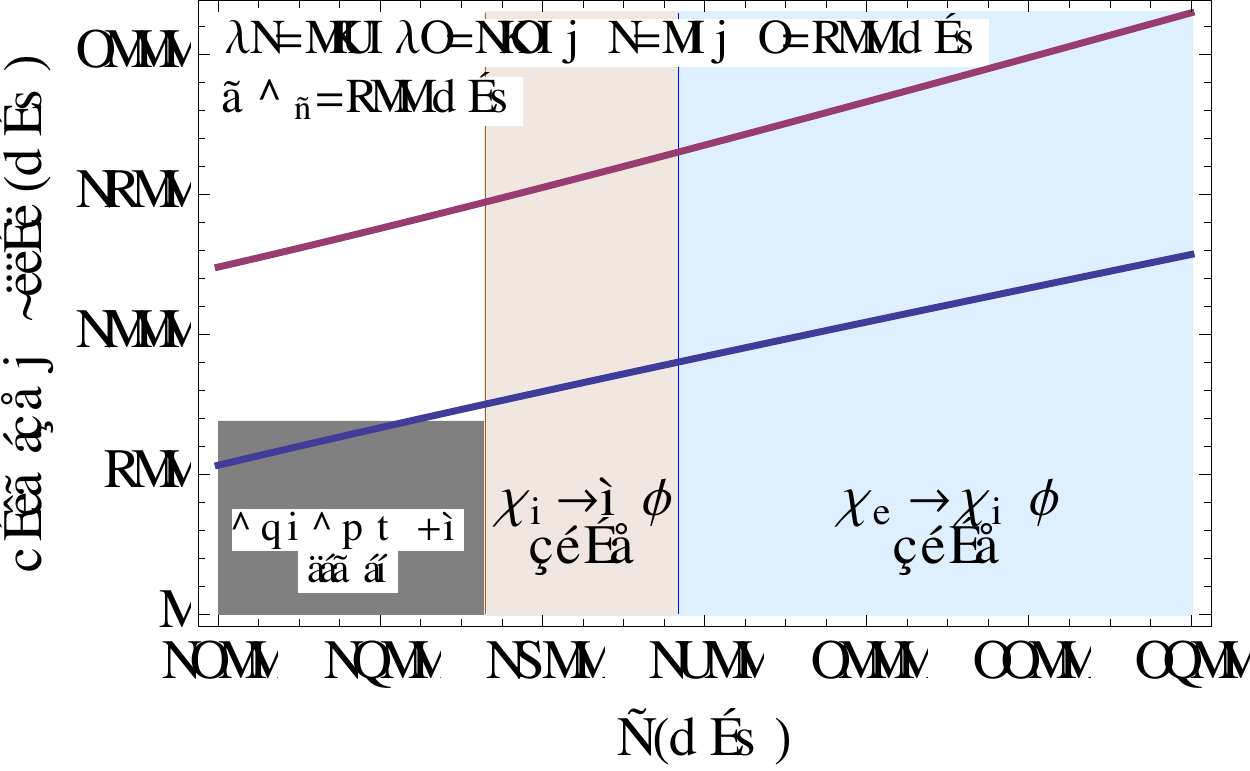}
\caption{\label{fig:fermionmasses} The masses of the degenerate charge $2/3$ and $5/3$ quarks in our model for the listed parameter values.  The shaded regions show where the decay modes $\chi_L \to u \phi$ and $\chi_H \to \chi_L \phi$ are kinematically open.}
\end{figure}

To start,  the masses of the fermions for our benchmark are plotted in Fig.~\ref{fig:fermionmasses} using Eq.~\ref{eq:limmixmass}.  Given large enough  $\lambda$ values, these quarks can be heavy enough to be safe from direct searches in the diphoton region of interest, but lower values of the couplings can be constrained.  The general phenomenology is as follows.  For the heavier quarks, as long as $M_1, M_2$ are comparable to $f$, there is substantial mixing in this sector, leading to cascades between the states if these modes are kinematically open.  For instance, the charge 5/3 and 2/3 $\chi_H$ quarks will decay into their partner quark $\chi_H\to \chi_L + \phi$ or $\to \chi_L + A_x$ if they are kinematically accessible.  As shown in the light blue shaded region of our benchmark, this occurs for large enough $f$ where the mass splitting is big enough to allow a $\phi$ particle to be emitted. 

The lighter quarks cannot decay until we consider the higher dimensional operator in Eq.~\ref{eq:couplings}.  The decays can be approximated by the equivalence principle
\bea
\chi_L^{2/3} &\to& u h\, (50\%),\; u Z\, (50\%), \nonumber \\ &\to& u \phi\, ((v/f)^2 50\%),\; u A_x\, ((v/f)^2 50\%)  
\\ \nonumber   \chi_L^{5/3} &\to& W^+ u\, (100\%)
\eea
where the approximate branching ratios are listed in parentheses and there is a implied weighted sum over decays into the various up-type quarks in the higher dimensional decay operator.  The $\phi$ and $A_x$ decays are suppressed by $(v/f)^2$ even when kinematically open (see light brown region in Fig.~\ref{fig:fermionmasses}), but could be interesting given a large sample of $\chi_L$ decays.    The most stringent LHC searches  constrain when the decays into  $u$ are just the top quark.  In that case, the charge $2/3$ quark is constrained to be above 855 GeV \cite{Aad:2015kqa,Khachatryan:2015oba} and the charge $5/3$ quark is constrained  to be heavier than 950 GeV \cite{CMSfivethirds}.  On the other hand, weaker limits exist when $u$ is just a light quark.  Here  an ATLAS search for quarks decaying to $W u$  constrains the $\chi_L^{5/3}$ mass to be larger than 690 GeV \cite{Aad:2015tba}.   That search also allows varying branching ratios into $q W, q Z, q h,$ but sets no limits for the expected branching ratios of the $\chi_L^{2/3}$.  Since the limit is so unconstraining for the charge 2/3 quark, a combination of the searches for both quarks would most likely lead to a similar limit.  In the general case with comparable branching ratios into top quarks and light quarks, these specific limits are weakened, but might lead to interesting mixed signals where a top quark and a light quark are produced.  As an idea of a limit which could apply to the most general ratio of decays into up-type quarks, we show the light quark limit of 690 GeV \cite{Aad:2015tba} in Fig.~\ref{fig:fermionmasses}.  As one can see, generalizing the decays of the new quarks into all type of up-type quarks opens up a larger part of the parameter space and leads to new mixed decay modes that are not currently being searched for.

\section{Conclusions \label{sec:conclusions}}
In this paper, we considered a simple theory explaining the ATLAS and CMS diphoton resonance excesses \cite{CMSATLAS13,CMS8,ATLAS8} that involves new exotic quarks charged under a new $U(1)_x$ gauge boson.  In this model, the diphoton resonance is the Higgs boson of the $U(1)_x$ theory $\phi$ and its signal rate to diphotons can be explained  given the large $5/3$ charges of the new quarks.  The additional $U(1)_x$ gauge boson was introduced as a motivation for the fundamental origin for $\phi$, but has the added benefit of  eliminating flavor constraints and leads to new correlated signals.  In particular,  new potential  signals for $\phi$  are decays into $\gamma A_x$ and $Z A_x$, with the $A_x$, through kinetic mixing, decaying back into Standard Model fermion pairs proportional to charge.  For the exotic quarks, the heavier quarks often decay into the lighter ones through emission of a $\phi$ or $A_x$ while the lighter quarks can decay into any of the up-type quarks since our model avoids flavor constraints.  

In conclusion, this model is a simple explanation of the diphoton excesses with nontrivial modifications to the most straightforward models with new exotic quarks.  In particular, the constraints on the quarks can be weakened by suppressed decays into top quarks.  At the same time, there are new signals involving the $\phi, A_x$ which can help to distinguish this from other diphoton explanations.    

\vspace{.1cm} \noindent \emph{Note Added:}  In the span of a week, as this project was starting and finishing up, a huge number of papers have appeared to explain the diphoton excess \cite{everybody}.    

\vspace{.1cm} \noindent \emph{Acknowledgements:} This work was supported in part by the Department of Energy under grants DE-SC0009945.  We thank G.~Kribs and B.~Ostdiek for discussions as well as early collaboration on an awesome, yet too strongly coupled model.   


\end{document}